\documentclass[reprint,superscriptaddress,amsmath,amssymb,aps,pre,longbibliography,floatfix]{revtex4-2}

\usepackage{amsmath}
\usepackage{amsfonts}
\usepackage{amssymb}
\usepackage{array}
\usepackage{bbm}
\usepackage{bm}
\usepackage{bbold}
\usepackage{braket}
\usepackage{bigints}
\usepackage{dsfont}
\usepackage[dvipsnames]{xcolor}
\usepackage{epsfig}
\usepackage{pdfpages}
\usepackage{tikz}

\usepackage{epstopdf}
\usepackage{pdfpages}
\usepackage{pgffor}
\usepackage{graphicx}
\usepackage{multirow}
\usepackage{mathtools}
\usepackage[T1]{fontenc}
\usepackage{newtxtext}
\usepackage[libertine]{newtxmath}
\usepackage{tensor}
\usepackage[utf8]{inputenc}
\usepackage[unicode=true,breaklinks=true,colorlinks=true]{hyperref}
\usepackage[all]{hypcap}
\usepackage{physics}

\hypersetup{citecolor=blue,urlcolor=blue}

\makeatletter
\AtBeginDocument{\let\LS@rot\@undefined}
\makeatother

\usepackage[normalem]{ulem}

\definecolor{cl1}{RGB}{0, 0, 0}
\definecolor{cl2}{RGB}{145, 82, 186}
\usepackage{ulem}

\begin{document}

\title{The rationality of radical pair mechanism in real biological systems}%

\author{Xiaoyu Chen}
\affiliation{School of Physics, Hubei Key Laboratory of Gravitation and Quantum Physics, Huazhong University of Science and Technology, Wuhan 430074, China}
\affiliation{International Joint Laboratory on Quantum Sensing and Quantum Metrology, Institute for Quantum Science and Engineering, Huazhong University of Science and Technology, Wuhan 430074, China}
\author{Haibin Liu}
\email{\textcolor{black}{Contact author: liuhb@hubu.edu.cn}}
\affiliation{School of Physics, Hubei University, Wuhan 430074, China}
\author{Jianming Cai}
\email{\textcolor{black}{Contact author: jianmingcai@hust.edu.cn}}
\affiliation{School of Physics, Hubei Key Laboratory of Gravitation and Quantum Physics, Huazhong University of Science and Technology, Wuhan 430074, China}
\affiliation{International Joint Laboratory on Quantum Sensing and Quantum Metrology, Institute for Quantum Science and Engineering, Huazhong University of Science and Technology, Wuhan 430074, China}

\date{\today}

\begin{abstract}  
The radical pair mechanism (RPM) in the chemical magnetic compass model is considered to be one of the most promising candidates for the avian magnetic navigation, and quantum needle phenomenon further boosts the navigation precision to a new high level. It is well known that there are also a variety of methods in the field of magnetic field sensing in laboratory, e.g. Ramsey protocol of NV centers in diamond. Here, we compare the RPM model and Ramsey-like model under laboratory conditions and under in vivo conditions respectively. The results are both surprising and reasonable. Under laboratory conditions, if we have precise control over time and a reasonably accurate prior knowledge of the magnetic field direction, the Ramsey-like model will outperform the RPM model. However, when such information is unavailable, as under in vivo conditions, the RPM model stands out. The RPM model achieves greater practicality at the cost of reduced accuracy.
\end{abstract}

\maketitle

\section{Introduction}
In recent years, the emerging field of quantum biology has attracted significant attention and research~\cite{ball2011physics,cai2016quantum,cao2020quantum,huelga2013vibrations,lambert2013quantum,mohseni2014quantum}, among which, one key research case is avian magnetoreception~\cite{wiltschko2006magnetoreception,maeda2008chemical}. The radical pair mechanism(RPM) has been repeatedly proven to be one of the strong candidates to explain the avian magnetic navigation in theoretical studies and behavioral experiments~\cite{bradlaugh2023essential,bassetto2023no,Jam_78_Sci,cai2010quantum,hogben2012entanglement,phillips1992behavioural,liu2017scheme,finkler2021quantum,aiq2022,wiltschko1972science,PhysRevLett.121.096001,PhysRevA.110.042220}. Relevant evidence suggests that photosensitive molecules of cryptochrome proteins form radical pairs under light excitation~\cite{xu2021magnetic,qin2016magnetic}, and may change their spin state under the influence of the magnetic field through the  radical pair mechanism, thereby changing the chemical reaction products and transmitting a magnetic signal~\cite{ritz2000model,rodgers2009chemical,ritz2004resonance}.

High precision has important biological significance in the study of animal spatial orientation, as migratory birds can use geomagnetic field information to navigate across continents with remarkable positioning accuracy~\cite{lefeldt2015migratory,chen2024identifying,aakesson2001avian,hiscock2016quantum}. Despite the existence of quantum effects such as spin coherence, quantum tunneling, and nuclear spin selectivity~\cite{cai2013chemical}, the performance of magnetic sensing in living organisms based on systems such as the radical pairs has not yet reached the quantum limit~\cite{PhysRevA.95.032129,smith2024optimality}, in stark contrast to artificial quantum sensing platforms. These platforms, such as nitrogen-vacancy centers, cold atom interferometers, can achieve measurement accuracy approaching standard quantum limit or even Heisenberg limit~\cite{degen2017quantum,schirhagl2014nitrogen,biedermann2015testing}, with precise control of various experimental parameters\cite{degen2017quantum}. Biological systems are inherently at a disadvantage, which may partly explain their relatively lower performance.  More specifically, from one perspective, biological systems face inherent environmental constraints: thermodynamic fluctuations at physiological temperature ($\sim300~\rm{K}$) cause quantum decoherence, molecular diffusion and collisions destroy quantum correlations, and environmental background noise further interferes with quantum effects~\cite{xiao2020magnetic,timmel1998effects}. From another, biological systems also do not have access to or control over the relevant experimental parameters.

In this study, we first compare a simplified RPM model that is constructed by a minimum two-level model with the classic quantum experimental paradigm of Ramsey-like interferometry~\cite{ramsey1949new,riehle1991optical,kaubruegger2021quantum}. Subsequently, we also introduce more realistic multi-spin radical pair to demonstrate the necessity for realizing biological functions and the rationality of its natural selection in comparison with Ramsey-like model.
Although the Ramsey-like method's multiple high-frequency oscillations achieve superior measurement accuracy to the single, high-precision signal provided by the quantum needle of radical pair, the presence of multiple oscillation peaks in the former places stringent demands on a prior knowledge of the magnetic field direction. After taking dephasing noise into account, the available time window for high-precision detection is limited to about 10 microseconds for Ramsey-like method, in the meanwhile RPM model stands out in the comparison, due to its long-term stability, provides a more convenient physical basis for the realization of magnetic navigation in birds.

Our work systematically evaluates the restrictions in artificial environments and the feasibility in the natural environment of the RPM and reveals the evolutionary rationality of this mechanism in biological magnetic induction. This cross-research perspective based on quantum physics and biology not only helps to clarify the physical limits of biological magnetic induction, but also provides an important theoretical basis for the optimal design of bionic quantum devices.

\section{Performance in highly controllable and simplified environment}
First, in order to more systematically explain the relevant physical mechanisms, we selected two typical two-level systems as research objects: one is the Ramsey-like interference system, and the other is a two-level quantum system constructed based on the multi-spin coupled radical pairs.

\subsection{Ramsey-like model and structural radical pair model}
In the Ramsey interferometry measurement~\cite{riehle1991optical,kaubruegger2021quantum}, a $\frac{\pi}{2}$ rotation around the $x$-axis is applied to the initialized two-level system firstly. In the second step, the system evolves for a period of time $t$ under the following Hamiltonian:
\begin{equation}
    H_{1}=\hbar\omega\sigma_z.
    \label{H1}
\end{equation}

For simplicity, the geomagnetic field is characterized by two parameters, magnetic field intensity $B$ and angle $\theta$, and $\omega$ in Eq. (\ref{H1}) is set as $\omega=Bcos\theta$, $\sigma_z$ is the Pauli matrix. 
Finally, another same $\frac{\pi}{2}$-pulse is implemented to the system, then the state is readout under as the expectation value of the projection operator $M=\ket{\psi_0}\bra{\psi_0}=\ket{0}\bra{0}$,
\begin{equation}
    F=\bra{\psi_f}M\ket{\psi_f}={\rm Tr} (\rho_{f} M),
    \label{ram-sol}
\end{equation}
where $\ket{\psi_f}=R_x(\frac{\pi}{2})e^{-\frac{i}{\hbar}H_{1}t}R_x(\frac{\pi}{2})\ket{\psi_0}$ is the final state of the evolution.

\begin{figure}
\includegraphics[width=1\columnwidth]{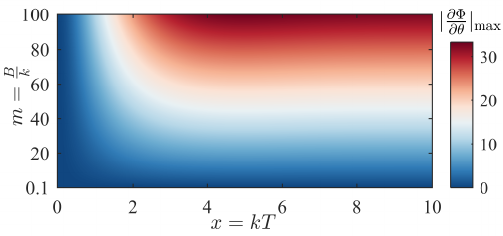}
\caption{\label{0820fig1} The maximum absolute value of the slope of product yield $\Phi$ for angle $\theta \in [0,\pi]$. }
\end{figure}

The radical pair model consists of two electron spins coupled by several nuclear spins~\cite{PhysRevA.85.040304,hore2016radical}. After light excitation, the electron spins are initially in a singlet state, which evolve under the influence of Zeeman interaction, hyperfine interaction, etc., resulting in coherent interconversion between the singlet state and the triplet state~\cite{hore2016radical,rodgers2009chemical}. Different directions of the geomagnetic field correspond to different product contents, thus providing navigation information for migratory birds\cite{kominis2009quantum,jones2010spin,hore2016radical}.

Similarly, we construct a two-level system with this structure and Hamiltonian is as following:
\begin{equation}
    H_{2}=\hbar\omega\sigma_x.
\end{equation}

To simplify complexity, the spin-selective recombination rates of $\ket{0}$ and $\ket{1}$ here are taken as the same value $k$. Given this scenario, we can employ an efficient method to solve the dynamics~\cite{doi:10.1080/00268979809483134}, which is described by
\begin{equation}
    \dot{\rho}=-\frac{i}{\hbar}[H_2 ,\rho]-k\rho.
    \label{rpm-ME}
\end{equation}
The solution to the Eq.~\eqref{rpm-ME} can be expressed in the form 
\begin{equation}
   \rho=e^{-\frac{i}{\hbar}H_2 t} \rho_0 e^{\frac{i}{\hbar}H_2 t} e^{-kt}.
\end{equation}
where $\rho_0=\ket{0}\bra{0}$ and the time-dependent probability of the system in $\ket{0}$ state is thus given by $p(t)={\rm Tr}[\rho(\ket{0}\bra{0})]$. 
After a period of time $T$, the product yield can be written as~\cite{doi:10.1080/00268979809483134}
 \begin{equation}
   \Phi=\int_{0}^{T}k p(t) dt.
\end{equation}

\subsection{Detection accuracy of magnetic filed angle }
In the study of avian navigation, the sensitivity of magnetic perception can be quantified by the partial derivative of the product yield with respect to the angle ($\abs{\partial \Phi/\partial \theta}$).
Similarly, in the Ramsey-like interferometer experiment, we also use the mathematical expression of the partial derivative of the expected value of the observed quantity with respect to the angle($\abs{\partial F/\partial \theta}$) to characterize the measurement accuracy of the system.

\subsubsection{An ideal condition}
Based on Eq.~\eqref{ram-sol}, the probability of observing the spin state in $\ket{0}$ is 
 \begin{equation}
 F=\frac{1-cos(\omega t)}{2}.
 \end{equation}
 Hence the accuracy in the estimated value of $\theta$ with Ramsey-like method is given by
 \begin{equation}
 \abs{\frac{\partial F}{\partial \theta}}=\frac{1}{2}\abs{sin(Btcos\theta)Btsin\theta}\leq \frac{\abs{Bt}}{2}.
 \end{equation}

Within the theoretical framework of the two-level model constructed based on the radical pair model, the fundamental object of our investigation is the product yield $\Phi$ which can be solved from time-dependent probability distribution $p(t)$. The formal proof proceeds as follows
\begin{equation}
p(t)=\frac{1+cos(\omega t)}{2}e^{-kt}.
\end{equation}
In light of the effect of recombination, we subsequently derive the expression for the product yield as follows
\begin{equation}
\begin{split}
   \Phi
    =&1-\frac{\omega^2}{2(k^2+\omega^2)}+\\
    &\frac{e^{-kT}\left[ -k^2-\omega^2-k^2cos(\omega T)+k\omega sin(\omega T)\right]}{2(k^2+\omega^2)},
\end{split}
\end{equation}
so the accuracy achieved can be quantified as
\begin{equation}
  \frac{\partial \Phi}{\partial \theta}=\frac{\partial \Phi}{\partial \omega}\cdot \frac{\partial \omega}{\partial \theta}=\frac{ \partial \Phi_{first}}{\partial \theta}+\frac{\partial \Phi_{second}}{\partial \theta}, 
\end{equation}
and
\begin{equation}
  \frac{\partial \Phi_{first}}{\partial \theta}
  =\frac{1}{(1/m_1+m_1)^2} \cdot tan\theta,
\end{equation}
\begin{equation}
\begin{split}
  \frac{\partial \Phi_{second}}{\partial \theta}
  =- \frac{e^{-x}tan\theta}{2(1/m_1+m_1)^2}\Big[(2+x+{m_1}^2 x)\cos(m_1 x)\\+
  (1/m_1+x/m_1-m_1+ m_1 x)\sin(m_1 x)\Big],
\end{split}
 \end{equation}
where $m_1=mcos \theta=\frac{B}{k}cos \theta$ and $x=kT$. The numerical simulation is shown in Fig.~\ref{0820fig1} about how $\abs{\frac{\partial \Phi}{\partial \theta}}_{\rm{max}}$ changes with $m$ and $x$. The result of $\abs{\frac{\partial \Phi}{\partial \theta}}_{\rm{max}} \leq \abs{\frac{\partial F}{\partial \theta}}_{\rm{max}}= \frac{\abs{Bt}}{2}=\frac{\abs{xm}}{2}$ is verified. And at a larger scale of $x$, the contribution of the second term to the slope of product yield is negligible compared to the first term, so $\abs{\frac{\partial \Phi}{\partial \theta}}_{\rm{max}}$ will not change much as $x$ increases.

\subsubsection{Markov noise}
Now we proceed to investigate this scenario under Markov noise, e.g. dephasing noise with a time-independent dephasing rate $\gamma$~\cite{gardiner2004quantum}. Incorporating decoherence, the time evolution of the density operator for a two-level system is governed by the following master equation~\cite{huelga1997improvement}:
\begin{equation}
    \dot{\rho}=-\frac{i}{\hbar}[H_1 ,\rho]+\frac{\gamma}{2}(\sigma_z \rho \sigma_z-\rho)
    \label{ram-me-n},
\end{equation}
which will enable the determination of the solution
\begin{equation}
 F=\frac{1-cos(\omega t)e^{-\gamma t}}{2},
 \end{equation}
 consequently, we obtain
\begin{equation}
\abs{\frac{\partial F}{\partial \theta}}=\frac{1}{2}\abs{e^{-\gamma t} sin(Btcos\theta)Btsin\theta}.
\label{eq:max_Ramsey}
 \end{equation}
 When $Btcos\theta=\frac{(2n+1)\pi}{2}$ and $t=\frac{1}{\gamma}$, the system attains its extremum, with the highest achievable accuracy being $\abs{\frac{B}{2e\gamma}}$.

For the constructed two-level model based on the radical pair mechanism, following the original model~\cite{haberkorn1976density}, we phenomenologically construct the master equation as follows

\begin{equation}
    \dot{\rho}=-\frac{i}{\hbar}[H_2 ,\rho]+\frac{\gamma}{2}(\sigma_x \rho \sigma_x-\rho)-\frac{k_1}{2}\{\rho,P_1\}-\frac{k_2}{2}\{\rho,P_2\},
    \label{eq:k1k2}
\end{equation}
so the corresponding yields under the two projection operators $P_1$ and $P_2$ are:
\begin{equation}
    \Phi_i=k_i \int_{0}^{\infty} {\rm Tr}(\rho P_i)dt.
\end{equation}
for two-level model $P_1=\ket{1}\bra{1}$, $P_2=\ket{0}\bra{0}$.
Firstly, We constrain ${k_1}^{-1}$ and ${k_2}^{-1}$ to $1-10~\rm{\mu s}$ to take into account the coherence time. In the absence of additional pure dephasing that is $\gamma=0$ in Eq.~\eqref{eq:k1k2}, when ${k_1}^{-1}={k_2}^{-1}=10~\rm{\mu s}$, the method can achieve the highest accuracy. This has been repeatedly verified: the longer the coherence lifetime, the better the performance of the chemical compass~\cite{hiscock2016quantum}.

\begin{figure}[b]
\includegraphics[width=1\columnwidth]{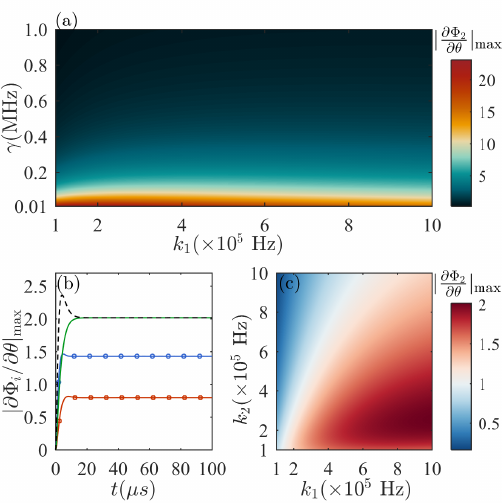}
\caption{\label{0820fig2}The maximum absolute value of the slope of product yield $\Phi_i$ for recombination rates ${k_i}\in [10^5,10^6]~\rm{Hz}$ and pure dephasing rate $\gamma \in [0.02,1.0]~{\rm MHz}(i=1,2)$. (a)Another recombination rate ${k_2}$ is set to be $10^5~\rm{Hz}$.
(b)The red solid line with squares(longest coherence time)  and the blue solid line with circles(equal recombination rate obtained at the highest accuracy) correspond to equal $k_i$ values of $10^5~\rm Hz$ and $5\times10^5~\rm Hz$ respectively. And the green solid line and the black dashed line(unequal recombination rates obtained at the highest accuracy) correspond to the changes of $\abs{\frac{\partial \Phi_2}{\partial \theta}}_{\rm{max}}$ and $\abs{\frac{\partial \Phi_1}{\partial \theta}}_{\rm{max}}$ with evolution time when $k_1=10\times10^5~\rm Hz$ and $k_2=2.6\times10^5~\rm Hz$, respectively. In addition, in (b) and (c), $\gamma$ is taken as $0.5~\rm{MHZ}$.}
\end{figure}

Given the recombination rates, it is reasonable that the larger the pure dephasing, the smaller the accuracy, as shown in Fig.~\ref{0820fig2}(a).
However, we also find that in the presence of a certain intensity of noise, the highest accuracy cannot be achieved when $k_1$ and $k_2$ take the same value. When the recombination rates are the same, the yields of the two products evolve synchronously over time. When the recombination rates are different, the yields of the two products change asynchronously for a period of time, but eventually stabilize and reach the same accuracy. In Fig.~\ref{0820fig2}(b), we select three sets of recombination rates shown in Fig.~\ref{0820fig2}(c) corresponding to the longest coherence time, the highest accuracy at equal recombination rates, and the highest accuracy at unequal recombination rates respectively. The results show that, unlike the non-noise case, it is not the longest coherence time that corresponds to the highest accuracy and that unequal recombination rates can achieve higher accuracy. However, it should be noted that the highest accuracy achievable under this situation is still lower than that of Eq.~\eqref{eq:max_Ramsey}.

\section{Performance in complex and realistic natural environments}
Generally speaking, some conditions that are taken for granted in the laboratory may seem to be very difficult for organisms in nature. For example, in the case of standard Ramsey spectroscopy, if we want to measure the maximum accuracy at a given time, we need to have control over the time at the microsecond level, and obtain a reasonably accurate prior knowledge of the parameter to be measured, and so on. This could be the very aspect that allows the RPM model to reverse its disadvantage.

\subsection{Theoretical model of RPM}
We started with the simplest model capable of producing the quantum needle that could reach the high precision required for songbirds' navigation~\cite{hiscock2016quantum,chen2024identifying}. The model consists of two radical electrons, each interacting with a nearby nucleus. The corresponding Hamiltonian is as follows~\cite{steiner1989magnetic}:
\begin{equation} \label{H}
H={I_1} \cdot \textbf{A}_1 \cdot {S_1} +{I_2} \cdot \textbf{A}_2 \cdot {S_2} +\gamma \vec{B} \cdot ({S_1}+{S_2}),
\end{equation}
where the system consists of two radical electronic spins, represented by the Pauli operators ${S_1}$ and ${S_2}$, each coupled to a nuclear spin, represented by the operators ${I_1}$ and ${I_2}$.
 The hyperfine interactions are characterized by tensors $\textbf{A}_1$ and $\textbf{A}_2$. The electron gyromagnetic ratio is defined as $\gamma=\frac{1}{2}\mu_0 g$, where $\mu_0$ denotes the Bohr magneton and  $g=2$ represents the electron Land\'{e} g-factor. We adopt the assumption that the separation of radical pair is large, rendering dipole-dipole and exchange interactions insignificant~\cite{o2005influence,efimova2008role}. The direction of the magnetic field is described by the angles $(\theta,\phi)$, with the field vector expressed as $\vec{B}=B_0(sin\theta cos\phi, sin\theta sin\phi, cos\theta)$, where $B_0$ is the field strength, and $\theta$ and $\phi$ represent the orientation of the magnetic field relative to the hyperfine tensor~\cite{PhysRevLett.106.040503}. Without loss of generality, we can set $\phi=0$ due to symmetry, restricting our analysis to $\theta$ in the range $[0,\pi]$.

The radical pair is assumed to be initially generated in a spin-correlated singlet electronic state through photoexcitation, while the relevant nuclear spins are in the maximally mixed state. Thus, the initial state can be expressed as $\rho_{0}=\frac{1}{N}P_{s}$, where $N$ counts all possible nuclear spin configurations. $P_s=\frac{1}{4}I-{S_1} \cdot {S_2}$ and $P_t=\frac{3}{4}I+{S_1} \cdot {S_2}$ project onto the singlet and triplet subspaces, respectively. The dynamic evolution of this system follows 
\begin{equation}
    \dot{\rho}=-\frac{i}{\hbar}[H,\rho]-\frac{k_s}{2}\{\rho,P_s\}-\frac{k_t}{2}\{\rho,P_t\}.
\end{equation}

\subsection{Mis-locking Avoidance}
\begin{figure}
\includegraphics[width=1\columnwidth]{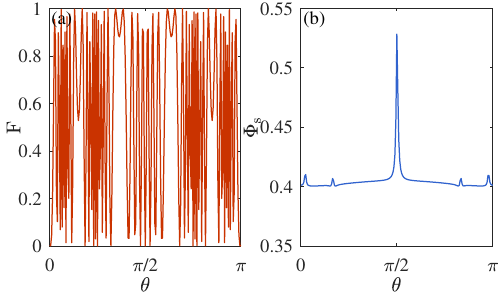}
\caption{\label{fig1} The variation of fidelity
$F$(Ramsey-like, left) and singlet yield $\Phi_s$(RPM, right) with angle $\theta \in [0,\pi]$. (b)The data $\textbf{A}_1=$diag[$0.0500$, $0.0006$, $0.0005$]~$\rm{mT}$, 
$\textbf{B}_1=$diag[$-0.0002$,$-0.0866$, $-0.0002$]~$\rm{mT}$, $M=R_{z}(-1.1465)R_{y}(-2.2840)R_{x}(1.2461)$, and $\textbf{A}_2=M\textbf{B}_1 M^T$ is taken from Ref.~\cite{chen2024identifying}.}
\end{figure}

Subsequently, we examine the implementation of the positioning mechanism in avian magnetoreception. As illustrated in Fig.~\ref{fig1}, the variations of the two signals as functions of the angle $\theta$ following an evolution period $T=10^{-4}~\rm{s}$ are presented. Additionally, it should be emphasized that for durations equal to or exceeding $T$, singlet yield signal shown in Fig.~\ref{fig1}(b) is almost stable. $B=50~{\mu \rm{T}}$ is selected to be close to the magnitude of the Earth's magnetic field and the radical pair lifetime is chosen as $k^{-1}=10$~${\mu \rm{s}}$.

As Fig.~\ref{fig1}(a) shown, $F$ oscillates multiple times as $\theta$ changes, which shows multiple similar peaks. Therefore, the prior condition that need to be known is that the range of the narrower $\theta$ which cannot include two oscillations of similar size or more in order to measure the value of the $\theta_{target}$. In contrast, there is only one sharp peak in Fig.~\ref{fig1}(b). As required by the navigation, no matter what global random initial conditions the migratory birds are in, they can always find the correct direction instead of falling into local optimality and getting lost.

\subsection{The influence of noise}
In both experimental and natural settings, environmental noise is an unavoidable factor that inevitably influences system dynamics. To rigorously characterize and compare the performance of the two models under study, we adopt a simplified noise model consisting solely of pure dephasing processes, as mathematically represented in Eq.~\eqref{ram-me-n}. Furthermore, the complete dynamical evolution of the radical pair model with four spins is governed by the Lindblad master equation~\cite{nielsen2010quantum,PhysRevLett.106.040503,PhysRevA.85.040304}:

\begin{eqnarray*}\label{eq:Lin}
\dot{\rho} & = & -\frac{i}{\hbar}\left[H,\rho\right]+\sum_{v}\frac{1}{2}\left(L_{v}\rho L_{v}^{\dagger}-\frac{1}{2}(L_{v}^{\dagger}L_{v}\rho+\rho L_{v}^{\dagger}L_{v})\right)\\
 &  & -\frac{k_{s}}{2}\{\rho,P_{s}\}-\frac{k_{t}}{2}\{\rho,P_{t}\}
\end{eqnarray*}
where ${L}_{1}$ = ${\gamma}^{1/2} \sigma_z^{(1)}$, $L_2$ = $\gamma^{1/2}\sigma_z^{(2)}$, $\gamma=\frac{1}{\tau_{dec}}$ and $\tau_{dec}$ is the decoherence time.

\begin{figure}
\includegraphics[width=1\columnwidth]{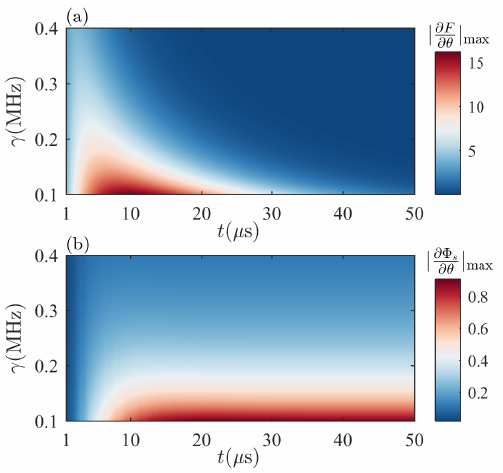}
\caption{\label{20250618fig3} Noise dependence of the absolute maximum slope in (a) Ramsey-like and (b) RPM models. Evolution time spans $1-50~\rm{\mu s}$, with dephasing rates $\gamma$ ranging from $0.1$ to $0.4~\rm{MHz}$. And the radical pair lifetime is chosen as $k^{-1}=10$~${\mu \rm{s}}$.}
\end{figure}

Under ideal conditions, the Ramsey-like model exhibits a linear proportionality between accuracy and evolution time. However, as Fig.~\ref{20250618fig3}(a) shown, in the presence of noise (for a given decoherence rate $\gamma$), the maximum absolute slope of fidelity in the Ramsey-like model initially increases, peaks, and subsequently decays to $0$. Notably, this peak occurs within a narrower temporal window, suggesting a trade-off between sensitivity and noise resilience.  In contrast, Fig.~\ref{20250618fig3}(b) demonstrates a distinct behavior of the RPM model: regardless of noise, the maximum absolute slope of the product yield first rises and then saturates at a near-constant value. A common trend in both models is that higher pure dephasing($\gamma$) systematically reduces the achievable accuracy.  

The Ramsey-like model requires careful selection of a short time window to observe significant slope values—a condition feasible in controlled experiments but impractical in natural biological environment. The RPM model, however, inherently supports such extended timescales due to its stable asymptotic behavior. Beyond $50~\rm{\mu s}$, the RPM maintains near-constant accuracy, while the Ramsey-like model’s performance degrades further. Thus, the RPM model offers superior accuracy on biologically relevant timescales, aligning with the requirements of continuous signal perception and response in living systems.  

\begin{figure}
\includegraphics[width=1\columnwidth]{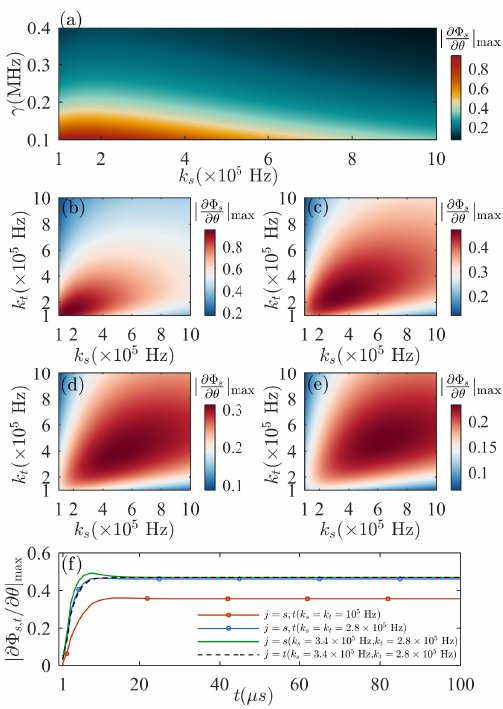}
\caption{\label{0812newfig2}The maximum absolute value of the slope of product yield $\Phi_{s,t}$ for recombination rates ${k_{s,t}}\in [10^5,10^6]~\rm{Hz}$ and pure dephasing rate $\gamma \in [0.1,0.4]~{\rm MHz}$. (a)Another recombination rate ${k_t}$ is set to be $10^5~\rm{Hz}$.
In (b)-(f), $\gamma$ is taken as $0.1, 0.2 ,0.3,0.4~\rm{MHZ}$. (f)The time evolution of the precision of singlet and triplet products for three sets of recombination rates(equal and minimum recombination rate, highest precision achieved with equal recombination rates, and highest precision achieved with unequal recombination rates) for given $\gamma=0.2~\rm{MHZ}$.}
\end{figure}

Taking different recombination rates into account, unequal recombination rates can still achieve higher accuracy as shown in Fig.~\ref{0812newfig2}. Furthermore, we also find that as $\gamma$ increases, the values of singlet and triplet recombination rates corresponding to high precision also gradually increase. This seems intelligent because higher noise intensity will shorten the lifetime of the radical pair. It seems that if migratory birds could adjust their own recombination rates to an adaptive range according to the noise intensity of the environment they are in, then magnetic navigation can achieve higher accuracy in the current environment.

\section{Discussion and conclusion}

In conclusion, this study aimed to evaluate and analyze the rationality of radical pair mechanism for navigation in real biological systems. By comparing the Ramsey-like method with RPM model both in highly controllable and simplified environment and in complex and realistic natural environments, we find that the Ramsey-like model performs better if with an environment where the parameters are highly controllable, otherwise RPM model that does not require precise time control and accurate prior knowledge of the measured parameters will stand out. It seems like that the evolutionary goal of magnetic navigation animals is not to pursue the highest theoretical accuracy, but to obtain navigation accuracy that meets survival needs while working with the least demanding conditions. Our research results not only provide a more comprehensive biological perspective on the existence of the radical pair mechanism in the magnetic navigation phenomenon, but also bring new insights into the construction of more universal quantum biosensors from a mechanistic level.

\section*{Acknowledgments}The computation was completed in the HPC Platform of Huazhong University of Science and Technology. This work is supported by the



\begin{thebibliography}{53}%
\makeatletter
\providecommand \@ifxundefined [1]{%
 \@ifx{#1\undefined}
}%
\providecommand \@ifnum [1]{%
 \ifnum #1\expandafter \@firstoftwo
 \else \expandafter \@secondoftwo
 \fi
}%
\providecommand \@ifx [1]{%
 \ifx #1\expandafter \@firstoftwo
 \else \expandafter \@secondoftwo
 \fi
}%
\providecommand \natexlab [1]{#1}%
\providecommand \enquote  [1]{``#1''}%
\providecommand \bibnamefont  [1]{#1}%
\providecommand \bibfnamefont [1]{#1}%
\providecommand \citenamefont [1]{#1}%
\providecommand \href@noop [0]{\@secondoftwo}%
\providecommand \href [0]{\begingroup \@sanitize@url \@href}%
\providecommand \@href[1]{\@@startlink{#1}\@@href}%
\providecommand \@@href[1]{\endgroup#1\@@endlink}%
\providecommand \@sanitize@url [0]{\catcode `\\12\catcode `\$12\catcode
  `\&12\catcode `\#12\catcode `\^12\catcode `\_12\catcode `\%12\relax}%
\providecommand \@@startlink[1]{}%
\providecommand \@@endlink[0]{}%
\providecommand \url  [0]{\begingroup\@sanitize@url \@url }%
\providecommand \@url [1]{\endgroup\@href {#1}{\urlprefix }}%
\providecommand \urlprefix  [0]{URL }%
\providecommand \Eprint [0]{\href }%
\providecommand \doibase [0]{https://doi.org/}%
\providecommand \selectlanguage [0]{\@gobble}%
\providecommand \bibinfo  [0]{\@secondoftwo}%
\providecommand \bibfield  [0]{\@secondoftwo}%
\providecommand \translation [1]{[#1]}%
\providecommand \BibitemOpen [0]{}%
\providecommand \bibitemStop [0]{}%
\providecommand \bibitemNoStop [0]{.\EOS\space}%
\providecommand \EOS [0]{\spacefactor3000\relax}%
\providecommand \BibitemShut  [1]{\csname bibitem#1\endcsname}%
\let\auto@bib@innerbib\@empty
\bibitem [{\citenamefont {Ball}(2011)}]{ball2011physics}%
  \BibitemOpen
  \bibfield  {author} {\bibinfo {author} {\bibfnamefont {P.}~\bibnamefont
  {Ball}},\ }\bibfield  {title} {\bibinfo {title} {{Physics of life: The dawn
  of quantum biology}},\ }\href
  {https://www.nature.com/articles/474272a#citeas} {\bibfield  {journal}
  {\bibinfo  {journal} {Nature}\ }\textbf {\bibinfo {volume} {474}},\ \bibinfo
  {pages} {272} (\bibinfo {year} {2011})}\BibitemShut {NoStop}%
\bibitem [{\citenamefont {Cai}(2016)}]{cai2016quantum}%
  \BibitemOpen
  \bibfield  {author} {\bibinfo {author} {\bibfnamefont {J.}~\bibnamefont
  {Cai}},\ }\bibfield  {title} {\bibinfo {title} {{Quantum biology: explore
  quantum dynamics in biological systems}},\ }\href
  {https://link.springer.com/article/10.1007/s11432-016-5592-y} {\bibfield
  {journal} {\bibinfo  {journal} {Sci. China Inf. Sci.}\ }\textbf {\bibinfo
  {volume} {59}},\ \bibinfo {pages} {1} (\bibinfo {year} {2016})}\BibitemShut
  {NoStop}%
\bibitem [{\citenamefont {Cao}\ \emph {et~al.}(2020)\citenamefont {Cao},
  \citenamefont {Cogdell}, \citenamefont {Coker}, \citenamefont {Duan},
  \citenamefont {Hauer}, \citenamefont {Kleinekath{\"o}fer}, \citenamefont
  {Jansen}, \citenamefont {Man{\v{c}}al}, \citenamefont {Miller}, \citenamefont
  {Ogilvie} \emph {et~al.}}]{cao2020quantum}%
  \BibitemOpen
  \bibfield  {author} {\bibinfo {author} {\bibfnamefont {J.}~\bibnamefont
  {Cao}}, \bibinfo {author} {\bibfnamefont {R.~J.}\ \bibnamefont {Cogdell}},
  \bibinfo {author} {\bibfnamefont {D.~F.}\ \bibnamefont {Coker}}, \bibinfo
  {author} {\bibfnamefont {H.-G.}\ \bibnamefont {Duan}}, \bibinfo {author}
  {\bibfnamefont {J.}~\bibnamefont {Hauer}}, \bibinfo {author} {\bibfnamefont
  {U.}~\bibnamefont {Kleinekath{\"o}fer}}, \bibinfo {author} {\bibfnamefont
  {T.~L.}\ \bibnamefont {Jansen}}, \bibinfo {author} {\bibfnamefont
  {T.}~\bibnamefont {Man{\v{c}}al}}, \bibinfo {author} {\bibfnamefont {R.~D.}\
  \bibnamefont {Miller}}, \bibinfo {author} {\bibfnamefont {J.~P.}\
  \bibnamefont {Ogilvie}}, \emph {et~al.},\ }\bibfield  {title} {\bibinfo
  {title} {{Quantum biology revisited}},\ }\href
  {https://www.science.org/doi/full/10.1126/sciadv.aaz4888} {\bibfield
  {journal} {\bibinfo  {journal} {Sci. Adv.}\ }\textbf {\bibinfo {volume}
  {6}},\ \bibinfo {pages} {eaaz4888} (\bibinfo {year} {2020})}\BibitemShut
  {NoStop}%
\bibitem [{\citenamefont {Huelga}\ and\ \citenamefont
  {Plenio}(2013)}]{huelga2013vibrations}%
  \BibitemOpen
  \bibfield  {author} {\bibinfo {author} {\bibfnamefont {S.~F.}\ \bibnamefont
  {Huelga}}\ and\ \bibinfo {author} {\bibfnamefont {M.~B.}\ \bibnamefont
  {Plenio}},\ }\bibfield  {title} {\bibinfo {title} {{Vibrations, quanta and
  biology}},\ }\href
  {https://www.tandfonline.com/doi/full/10.1080/00405000.2013.829687}
  {\bibfield  {journal} {\bibinfo  {journal} {Contemp. Phys.}\ }\textbf
  {\bibinfo {volume} {54}},\ \bibinfo {pages} {181} (\bibinfo {year}
  {2013})}\BibitemShut {NoStop}%
\bibitem [{\citenamefont {Lambert}\ \emph {et~al.}(2013)\citenamefont
  {Lambert}, \citenamefont {Chen}, \citenamefont {Cheng}, \citenamefont {Li},
  \citenamefont {Chen},\ and\ \citenamefont {Nori}}]{lambert2013quantum}%
  \BibitemOpen
  \bibfield  {author} {\bibinfo {author} {\bibfnamefont {N.}~\bibnamefont
  {Lambert}}, \bibinfo {author} {\bibfnamefont {Y.-N.}\ \bibnamefont {Chen}},
  \bibinfo {author} {\bibfnamefont {Y.-C.}\ \bibnamefont {Cheng}}, \bibinfo
  {author} {\bibfnamefont {C.-M.}\ \bibnamefont {Li}}, \bibinfo {author}
  {\bibfnamefont {G.-Y.}\ \bibnamefont {Chen}},\ and\ \bibinfo {author}
  {\bibfnamefont {F.}~\bibnamefont {Nori}},\ }\bibfield  {title} {\bibinfo
  {title} {{Quantum biology}},\ }\href
  {https://www.nature.com/articles/nphys2474} {\bibfield  {journal} {\bibinfo
  {journal} {Nat. Phys.}\ }\textbf {\bibinfo {volume} {9}},\ \bibinfo {pages}
  {10} (\bibinfo {year} {2013})}\BibitemShut {NoStop}%
\bibitem [{\citenamefont {Mohseni}\ \emph {et~al.}(2014)\citenamefont
  {Mohseni}, \citenamefont {Omar}, \citenamefont {Engel},\ and\ \citenamefont
  {Plenio}}]{mohseni2014quantum}%
  \BibitemOpen
  \bibfield  {author} {\bibinfo {author} {\bibfnamefont {M.}~\bibnamefont
  {Mohseni}}, \bibinfo {author} {\bibfnamefont {Y.}~\bibnamefont {Omar}},
  \bibinfo {author} {\bibfnamefont {G.~S.}\ \bibnamefont {Engel}},\ and\
  \bibinfo {author} {\bibfnamefont {M.~B.}\ \bibnamefont {Plenio}},\
  }\href@noop {} {\emph {\bibinfo {title} {{Quantum effects in biology}}}}\
  (\bibinfo  {publisher} {Cambridge University Press, Cambridge},\ \bibinfo
  {year} {2014})\BibitemShut {NoStop}%
\bibitem [{\citenamefont {Wiltschko}\ and\ \citenamefont
  {Wiltschko}(2006)}]{wiltschko2006magnetoreception}%
  \BibitemOpen
  \bibfield  {author} {\bibinfo {author} {\bibfnamefont {R.}~\bibnamefont
  {Wiltschko}}\ and\ \bibinfo {author} {\bibfnamefont {W.}~\bibnamefont
  {Wiltschko}},\ }\bibfield  {title} {\bibinfo {title} {{Magnetoreception}},\
  }\href {https://onlinelibrary.wiley.com/doi/abs/10.1002/bies.20363}
  {\bibfield  {journal} {\bibinfo  {journal} {BioEssays}\ }\textbf {\bibinfo
  {volume} {28}},\ \bibinfo {pages} {157} (\bibinfo {year} {2006})}\BibitemShut
  {NoStop}%
\bibitem [{\citenamefont {Maeda}\ \emph {et~al.}(2008)\citenamefont {Maeda},
  \citenamefont {Henbest}, \citenamefont {Cintolesi}, \citenamefont {Kuprov},
  \citenamefont {Rodgers}, \citenamefont {Liddell}, \citenamefont {Gust},
  \citenamefont {Timmel},\ and\ \citenamefont {Hore}}]{maeda2008chemical}%
  \BibitemOpen
  \bibfield  {author} {\bibinfo {author} {\bibfnamefont {K.}~\bibnamefont
  {Maeda}}, \bibinfo {author} {\bibfnamefont {K.~B.}\ \bibnamefont {Henbest}},
  \bibinfo {author} {\bibfnamefont {F.}~\bibnamefont {Cintolesi}}, \bibinfo
  {author} {\bibfnamefont {I.}~\bibnamefont {Kuprov}}, \bibinfo {author}
  {\bibfnamefont {C.~T.}\ \bibnamefont {Rodgers}}, \bibinfo {author}
  {\bibfnamefont {P.~A.}\ \bibnamefont {Liddell}}, \bibinfo {author}
  {\bibfnamefont {D.}~\bibnamefont {Gust}}, \bibinfo {author} {\bibfnamefont
  {C.~R.}\ \bibnamefont {Timmel}},\ and\ \bibinfo {author} {\bibfnamefont
  {P.~J.}\ \bibnamefont {Hore}},\ }\bibfield  {title} {\bibinfo {title}
  {{Chemical compass model of avian magnetoreception}},\ }\href
  {https://www.nature.com/articles/nature06834} {\bibfield  {journal} {\bibinfo
   {journal} {Nature}\ }\textbf {\bibinfo {volume} {453}},\ \bibinfo {pages}
  {387} (\bibinfo {year} {2008})}\BibitemShut {NoStop}%
\bibitem [{\citenamefont {Bradlaugh}\ \emph {et~al.}(2023)\citenamefont
  {Bradlaugh}, \citenamefont {Fedele}, \citenamefont {Munro}, \citenamefont
  {Hansen}, \citenamefont {Hares}, \citenamefont {Patel}, \citenamefont
  {Kyriacou}, \citenamefont {Jones}, \citenamefont {Rosato},\ and\
  \citenamefont {Baines}}]{bradlaugh2023essential}%
  \BibitemOpen
  \bibfield  {author} {\bibinfo {author} {\bibfnamefont {A.~A.}\ \bibnamefont
  {Bradlaugh}}, \bibinfo {author} {\bibfnamefont {G.}~\bibnamefont {Fedele}},
  \bibinfo {author} {\bibfnamefont {A.~L.}\ \bibnamefont {Munro}}, \bibinfo
  {author} {\bibfnamefont {C.~N.}\ \bibnamefont {Hansen}}, \bibinfo {author}
  {\bibfnamefont {J.~M.}\ \bibnamefont {Hares}}, \bibinfo {author}
  {\bibfnamefont {S.}~\bibnamefont {Patel}}, \bibinfo {author} {\bibfnamefont
  {C.~P.}\ \bibnamefont {Kyriacou}}, \bibinfo {author} {\bibfnamefont {A.~R.}\
  \bibnamefont {Jones}}, \bibinfo {author} {\bibfnamefont {E.}~\bibnamefont
  {Rosato}},\ and\ \bibinfo {author} {\bibfnamefont {R.~A.}\ \bibnamefont
  {Baines}},\ }\bibfield  {title} {\bibinfo {title} {{Essential elements of
  radical pair magnetosensitivity in Drosophila}},\ }\href
  {https://www.nature.com/articles/s41586-023-05735-z} {\bibfield  {journal}
  {\bibinfo  {journal} {Nature}\ }\textbf {\bibinfo {volume} {615}},\ \bibinfo
  {pages} {111} (\bibinfo {year} {2023})}\BibitemShut {NoStop}%
\bibitem [{\citenamefont {Bassetto}\ \emph {et~al.}(2023)\citenamefont
  {Bassetto}, \citenamefont {Reichl}, \citenamefont {Kobylkov}, \citenamefont
  {Kattnig}, \citenamefont {Winklhofer}, \citenamefont {Hore},\ and\
  \citenamefont {Mouritsen}}]{bassetto2023no}%
  \BibitemOpen
  \bibfield  {author} {\bibinfo {author} {\bibfnamefont {M.}~\bibnamefont
  {Bassetto}}, \bibinfo {author} {\bibfnamefont {T.}~\bibnamefont {Reichl}},
  \bibinfo {author} {\bibfnamefont {D.}~\bibnamefont {Kobylkov}}, \bibinfo
  {author} {\bibfnamefont {D.~R.}\ \bibnamefont {Kattnig}}, \bibinfo {author}
  {\bibfnamefont {M.}~\bibnamefont {Winklhofer}}, \bibinfo {author}
  {\bibfnamefont {P.}~\bibnamefont {Hore}},\ and\ \bibinfo {author}
  {\bibfnamefont {H.}~\bibnamefont {Mouritsen}},\ }\bibfield  {title} {\bibinfo
  {title} {{No evidence for magnetic field effects on the behaviour of
  Drosophila}},\ }\href {https://www.nature.com/articles/s41586-023-06397-7}
  {\bibfield  {journal} {\bibinfo  {journal} {Nature}\ }\textbf {\bibinfo
  {volume} {620}},\ \bibinfo {pages} {595} (\bibinfo {year}
  {2023})}\BibitemShut {NoStop}%
\bibitem [{\citenamefont {Gould}\ \emph {et~al.}(1978)\citenamefont {Gould},
  \citenamefont {Kirschvink},\ and\ \citenamefont {Deffeyes}}]{Jam_78_Sci}%
  \BibitemOpen
  \bibfield  {author} {\bibinfo {author} {\bibfnamefont {J.~L.}\ \bibnamefont
  {Gould}}, \bibinfo {author} {\bibfnamefont {J.~L.}\ \bibnamefont
  {Kirschvink}},\ and\ \bibinfo {author} {\bibfnamefont {K.~S.}\ \bibnamefont
  {Deffeyes}},\ }\bibfield  {title} {\bibinfo {title} {{Bees Have Magnetic
  Remanence}},\ }\href
  {https://www.science.org/doi/abs/10.1126/science.201.4360.1026} {\bibfield
  {journal} {\bibinfo  {journal} {Science}\ }\textbf {\bibinfo {volume}
  {201}},\ \bibinfo {pages} {1026} (\bibinfo {year} {1978})}\BibitemShut
  {NoStop}%
\bibitem [{\citenamefont {Cai}\ \emph {et~al.}(2010)\citenamefont {Cai},
  \citenamefont {Guerreschi},\ and\ \citenamefont {Briegel}}]{cai2010quantum}%
  \BibitemOpen
  \bibfield  {author} {\bibinfo {author} {\bibfnamefont {J.}~\bibnamefont
  {Cai}}, \bibinfo {author} {\bibfnamefont {G.~G.}\ \bibnamefont
  {Guerreschi}},\ and\ \bibinfo {author} {\bibfnamefont {H.~J.}\ \bibnamefont
  {Briegel}},\ }\bibfield  {title} {\bibinfo {title} {{Quantum control and
  entanglement in a chemical compass}},\ }\href
  {https://journals.aps.org/prl/abstract/10.1103/PhysRevLett.104.220502}
  {\bibfield  {journal} {\bibinfo  {journal} {Phys. Rev. Lett.}\ }\textbf
  {\bibinfo {volume} {104}},\ \bibinfo {pages} {220502} (\bibinfo {year}
  {2010})}\BibitemShut {NoStop}%
\bibitem [{\citenamefont {Hogben}\ \emph {et~al.}(2012)\citenamefont {Hogben},
  \citenamefont {Biskup},\ and\ \citenamefont {Hore}}]{hogben2012entanglement}%
  \BibitemOpen
  \bibfield  {author} {\bibinfo {author} {\bibfnamefont {H.~J.}\ \bibnamefont
  {Hogben}}, \bibinfo {author} {\bibfnamefont {T.}~\bibnamefont {Biskup}},\
  and\ \bibinfo {author} {\bibfnamefont {P.}~\bibnamefont {Hore}},\ }\bibfield
  {title} {\bibinfo {title} {{Entanglement and sources of magnetic anisotropy
  in radical pair-based avian magnetoreceptors}},\ }\href
  {https://journals.aps.org/prl/abstract/10.1103/PhysRevLett.109.220501}
  {\bibfield  {journal} {\bibinfo  {journal} {Phys. Rev. Lett.}\ }\textbf
  {\bibinfo {volume} {109}},\ \bibinfo {pages} {220501} (\bibinfo {year}
  {2012})}\BibitemShut {NoStop}%
\bibitem [{\citenamefont {Phillips}\ and\ \citenamefont
  {Borland}(1992)}]{phillips1992behavioural}%
  \BibitemOpen
  \bibfield  {author} {\bibinfo {author} {\bibfnamefont {J.~B.}\ \bibnamefont
  {Phillips}}\ and\ \bibinfo {author} {\bibfnamefont {S.~C.}\ \bibnamefont
  {Borland}},\ }\bibfield  {title} {\bibinfo {title} {{Behavioural evidence for
  use of a light-dependent magnetoreception mechanism by a vertebrate}},\
  }\href {https://www.nature.com/articles/359142a0} {\bibfield  {journal}
  {\bibinfo  {journal} {Nature}\ }\textbf {\bibinfo {volume} {359}},\ \bibinfo
  {pages} {142} (\bibinfo {year} {1992})}\BibitemShut {NoStop}%
\bibitem [{\citenamefont {Liu}\ \emph {et~al.}(2017)\citenamefont {Liu},
  \citenamefont {Plenio},\ and\ \citenamefont {Cai}}]{liu2017scheme}%
  \BibitemOpen
  \bibfield  {author} {\bibinfo {author} {\bibfnamefont {H.}~\bibnamefont
  {Liu}}, \bibinfo {author} {\bibfnamefont {M.~B.}\ \bibnamefont {Plenio}},\
  and\ \bibinfo {author} {\bibfnamefont {J.}~\bibnamefont {Cai}},\ }\bibfield
  {title} {\bibinfo {title} {{Scheme for detection of single-molecule radical
  pair reaction using spin in diamond}},\ }\href
  {https://journals.aps.org/prl/abstract/10.1103/PhysRevLett.118.200402}
  {\bibfield  {journal} {\bibinfo  {journal} {Phys. Rev. Lett.}\ }\textbf
  {\bibinfo {volume} {118}},\ \bibinfo {pages} {200402} (\bibinfo {year}
  {2017})}\BibitemShut {NoStop}%
\bibitem [{\citenamefont {Finkler}\ and\ \citenamefont
  {Dasari}(2021)}]{finkler2021quantum}%
  \BibitemOpen
  \bibfield  {author} {\bibinfo {author} {\bibfnamefont {A.}~\bibnamefont
  {Finkler}}\ and\ \bibinfo {author} {\bibfnamefont {D.}~\bibnamefont
  {Dasari}},\ }\bibfield  {title} {\bibinfo {title} {{Quantum sensing and
  control of spin-state dynamics in the radical-pair mechanism}},\ }\href
  {https://journals.aps.org/prapplied/abstract/10.1103/PhysRevApplied.15.034066}
  {\bibfield  {journal} {\bibinfo  {journal} {Phys. Rev. Appl.}\ }\textbf
  {\bibinfo {volume} {15}},\ \bibinfo {pages} {034066} (\bibinfo {year}
  {2021})}\BibitemShut {NoStop}%
\bibitem [{\citenamefont {Wu}\ \emph {et~al.}(2022)\citenamefont {Wu},
  \citenamefont {Hu}, \citenamefont {Zhu}, \citenamefont {Deng},\ and\
  \citenamefont {Ai}}]{aiq2022}%
  \BibitemOpen
  \bibfield  {author} {\bibinfo {author} {\bibfnamefont {J.-Y.}\ \bibnamefont
  {Wu}}, \bibinfo {author} {\bibfnamefont {X.-Y.}\ \bibnamefont {Hu}}, \bibinfo
  {author} {\bibfnamefont {H.-Y.}\ \bibnamefont {Zhu}}, \bibinfo {author}
  {\bibfnamefont {R.-Q.}\ \bibnamefont {Deng}},\ and\ \bibinfo {author}
  {\bibfnamefont {Q.}~\bibnamefont {Ai}},\ }\bibfield  {title} {\bibinfo
  {title} {{A Bionic Compass Based on Multiradicals}},\ }\href
  {https://doi.org/10.1021/acs.jpcb.2c02711} {\bibfield  {journal} {\bibinfo
  {journal} {J. Phys. Chem. B}\ }\textbf {\bibinfo {volume} {126}},\ \bibinfo
  {pages} {10327} (\bibinfo {year} {2022})}\BibitemShut {NoStop}%
\bibitem [{\citenamefont {Wiltschko}\ and\ \citenamefont
  {Wiltschko}(1972)}]{wiltschko1972science}%
  \BibitemOpen
  \bibfield  {author} {\bibinfo {author} {\bibfnamefont {W.}~\bibnamefont
  {Wiltschko}}\ and\ \bibinfo {author} {\bibfnamefont {R.}~\bibnamefont
  {Wiltschko}},\ }\bibfield  {title} {\bibinfo {title} {{Magnetic Compass of
  European Robins}},\ }\href
  {https://www.science.org/doi/abs/10.1126/science.176.4030.62} {\bibfield
  {journal} {\bibinfo  {journal} {Science}\ }\textbf {\bibinfo {volume}
  {176}},\ \bibinfo {pages} {62} (\bibinfo {year} {1972})}\BibitemShut
  {NoStop}%
\bibitem [{\citenamefont {Keens}\ \emph {et~al.}(2018)\citenamefont {Keens},
  \citenamefont {Bedkihal},\ and\ \citenamefont
  {Kattnig}}]{PhysRevLett.121.096001}%
  \BibitemOpen
  \bibfield  {author} {\bibinfo {author} {\bibfnamefont {R.~H.}\ \bibnamefont
  {Keens}}, \bibinfo {author} {\bibfnamefont {S.}~\bibnamefont {Bedkihal}},\
  and\ \bibinfo {author} {\bibfnamefont {D.~R.}\ \bibnamefont {Kattnig}},\
  }\bibfield  {title} {\bibinfo {title} {{Magnetosensitivity in Dipolarly
  Coupled Three-Spin Systems}},\ }\href
  {https://doi.org/10.1103/PhysRevLett.121.096001} {\bibfield  {journal}
  {\bibinfo  {journal} {Phys. Rev. Lett.}\ }\textbf {\bibinfo {volume} {121}},\
  \bibinfo {pages} {096001} (\bibinfo {year} {2018})}\BibitemShut {NoStop}%
\bibitem [{\citenamefont {Chen}\ \emph
  {et~al.}(2024{\natexlab{a}})\citenamefont {Chen}, \citenamefont {Liu},\ and\
  \citenamefont {Cai}}]{PhysRevA.110.042220}%
  \BibitemOpen
  \bibfield  {author} {\bibinfo {author} {\bibfnamefont {X.}~\bibnamefont
  {Chen}}, \bibinfo {author} {\bibfnamefont {H.}~\bibnamefont {Liu}},\ and\
  \bibinfo {author} {\bibfnamefont {J.}~\bibnamefont {Cai}},\ }\bibfield
  {title} {\bibinfo {title} {Identifying a possible mechanism for the quantum
  needle in chemical magnetoreception},\ }\href
  {https://doi.org/10.1103/PhysRevA.110.042220} {\bibfield  {journal} {\bibinfo
   {journal} {Phys. Rev. A}\ }\textbf {\bibinfo {volume} {110}},\ \bibinfo
  {pages} {042220} (\bibinfo {year} {2024}{\natexlab{a}})}\BibitemShut
  {NoStop}%
\bibitem [{\citenamefont {Xu}\ \emph {et~al.}(2021)\citenamefont {Xu},
  \citenamefont {Jarocha}, \citenamefont {Zollitsch}, \citenamefont
  {Konowalczyk}, \citenamefont {Henbest}, \citenamefont {Richert},
  \citenamefont {Golesworthy}, \citenamefont {Schmidt}, \citenamefont
  {D{\'e}jean}, \citenamefont {Sowood} \emph {et~al.}}]{xu2021magnetic}%
  \BibitemOpen
  \bibfield  {author} {\bibinfo {author} {\bibfnamefont {J.}~\bibnamefont
  {Xu}}, \bibinfo {author} {\bibfnamefont {L.~E.}\ \bibnamefont {Jarocha}},
  \bibinfo {author} {\bibfnamefont {T.}~\bibnamefont {Zollitsch}}, \bibinfo
  {author} {\bibfnamefont {M.}~\bibnamefont {Konowalczyk}}, \bibinfo {author}
  {\bibfnamefont {K.~B.}\ \bibnamefont {Henbest}}, \bibinfo {author}
  {\bibfnamefont {S.}~\bibnamefont {Richert}}, \bibinfo {author} {\bibfnamefont
  {M.~J.}\ \bibnamefont {Golesworthy}}, \bibinfo {author} {\bibfnamefont
  {J.}~\bibnamefont {Schmidt}}, \bibinfo {author} {\bibfnamefont
  {V.}~\bibnamefont {D{\'e}jean}}, \bibinfo {author} {\bibfnamefont {D.~J.}\
  \bibnamefont {Sowood}}, \emph {et~al.},\ }\bibfield  {title} {\bibinfo
  {title} {{Magnetic sensitivity of cryptochrome 4 from a migratory
  songbird}},\ }\href {https://www.nature.com/articles/s41586-021-03618-9}
  {\bibfield  {journal} {\bibinfo  {journal} {Nature}\ }\textbf {\bibinfo
  {volume} {594}},\ \bibinfo {pages} {535} (\bibinfo {year}
  {2021})}\BibitemShut {NoStop}%
\bibitem [{\citenamefont {Qin}\ \emph {et~al.}(2016)\citenamefont {Qin},
  \citenamefont {Yin}, \citenamefont {Yang}, \citenamefont {Dou}, \citenamefont
  {Liu}, \citenamefont {Zhang}, \citenamefont {Yu}, \citenamefont {Huang},
  \citenamefont {Feng}, \citenamefont {Hao} \emph {et~al.}}]{qin2016magnetic}%
  \BibitemOpen
  \bibfield  {author} {\bibinfo {author} {\bibfnamefont {S.}~\bibnamefont
  {Qin}}, \bibinfo {author} {\bibfnamefont {H.}~\bibnamefont {Yin}}, \bibinfo
  {author} {\bibfnamefont {C.}~\bibnamefont {Yang}}, \bibinfo {author}
  {\bibfnamefont {Y.}~\bibnamefont {Dou}}, \bibinfo {author} {\bibfnamefont
  {Z.}~\bibnamefont {Liu}}, \bibinfo {author} {\bibfnamefont {P.}~\bibnamefont
  {Zhang}}, \bibinfo {author} {\bibfnamefont {H.}~\bibnamefont {Yu}}, \bibinfo
  {author} {\bibfnamefont {Y.}~\bibnamefont {Huang}}, \bibinfo {author}
  {\bibfnamefont {J.}~\bibnamefont {Feng}}, \bibinfo {author} {\bibfnamefont
  {J.}~\bibnamefont {Hao}}, \emph {et~al.},\ }\bibfield  {title} {\bibinfo
  {title} {{A magnetic protein biocompass}},\ }\href
  {https://www.nature.com/articles/nmat4484} {\bibfield  {journal} {\bibinfo
  {journal} {Nat. Mater.}\ }\textbf {\bibinfo {volume} {15}},\ \bibinfo {pages}
  {217} (\bibinfo {year} {2016})}\BibitemShut {NoStop}%
\bibitem [{\citenamefont {Ritz}\ \emph {et~al.}(2000)\citenamefont {Ritz},
  \citenamefont {Adem},\ and\ \citenamefont {Schulten}}]{ritz2000model}%
  \BibitemOpen
  \bibfield  {author} {\bibinfo {author} {\bibfnamefont {T.}~\bibnamefont
  {Ritz}}, \bibinfo {author} {\bibfnamefont {S.}~\bibnamefont {Adem}},\ and\
  \bibinfo {author} {\bibfnamefont {K.}~\bibnamefont {Schulten}},\ }\bibfield
  {title} {\bibinfo {title} {{A model for photoreceptor-based magnetoreception
  in birds}},\ }\href {https://www.cell.com/fulltext/S0006-3495%2800%2976629-X}
  {\bibfield  {journal} {\bibinfo  {journal} {Biophys. J.}\ }\textbf {\bibinfo
  {volume} {78}},\ \bibinfo {pages} {707} (\bibinfo {year} {2000})}\BibitemShut
  {NoStop}%
\bibitem [{\citenamefont {Rodgers}\ and\ \citenamefont
  {Hore}(2009)}]{rodgers2009chemical}%
  \BibitemOpen
  \bibfield  {author} {\bibinfo {author} {\bibfnamefont {C.~T.}\ \bibnamefont
  {Rodgers}}\ and\ \bibinfo {author} {\bibfnamefont {P.~J.}\ \bibnamefont
  {Hore}},\ }\bibfield  {title} {\bibinfo {title} {{Chemical magnetoreception
  in birds: the radical pair mechanism}},\ }\href
  {https://www.pnas.org/doi/abs/10.1073/pnas.0711968106} {\bibfield  {journal}
  {\bibinfo  {journal} {Proc. Natl. Acad. Sci. USA}\ }\textbf {\bibinfo
  {volume} {106}},\ \bibinfo {pages} {353} (\bibinfo {year}
  {2009})}\BibitemShut {NoStop}%
\bibitem [{\citenamefont {Ritz}\ \emph {et~al.}(2004)\citenamefont {Ritz},
  \citenamefont {Thalau}, \citenamefont {Phillips}, \citenamefont {Wiltschko},\
  and\ \citenamefont {Wiltschko}}]{ritz2004resonance}%
  \BibitemOpen
  \bibfield  {author} {\bibinfo {author} {\bibfnamefont {T.}~\bibnamefont
  {Ritz}}, \bibinfo {author} {\bibfnamefont {P.}~\bibnamefont {Thalau}},
  \bibinfo {author} {\bibfnamefont {J.~B.}\ \bibnamefont {Phillips}}, \bibinfo
  {author} {\bibfnamefont {R.}~\bibnamefont {Wiltschko}},\ and\ \bibinfo
  {author} {\bibfnamefont {W.}~\bibnamefont {Wiltschko}},\ }\bibfield  {title}
  {\bibinfo {title} {{Resonance effects indicate a radical-pair mechanism for
  avian magnetic compass}},\ }\href
  {https://www.nature.com/articles/nature02534} {\bibfield  {journal} {\bibinfo
   {journal} {Nature}\ }\textbf {\bibinfo {volume} {429}},\ \bibinfo {pages}
  {177} (\bibinfo {year} {2004})}\BibitemShut {NoStop}%
\bibitem [{\citenamefont {Lefeldt}\ \emph {et~al.}(2015)\citenamefont
  {Lefeldt}, \citenamefont {Dreyer}, \citenamefont {Schneider}, \citenamefont
  {Steenken},\ and\ \citenamefont {Mouritsen}}]{lefeldt2015migratory}%
  \BibitemOpen
  \bibfield  {author} {\bibinfo {author} {\bibfnamefont {N.}~\bibnamefont
  {Lefeldt}}, \bibinfo {author} {\bibfnamefont {D.}~\bibnamefont {Dreyer}},
  \bibinfo {author} {\bibfnamefont {N.-L.}\ \bibnamefont {Schneider}}, \bibinfo
  {author} {\bibfnamefont {F.}~\bibnamefont {Steenken}},\ and\ \bibinfo
  {author} {\bibfnamefont {H.}~\bibnamefont {Mouritsen}},\ }\bibfield  {title}
  {\bibinfo {title} {{Migratory blackcaps tested in Emlen funnels can orient at
  85 degrees but not at 88 degrees magnetic inclination}},\ }\href
  {https://journals.biologists.com/jeb/article/218/2/206/14271/Migratory-blackcaps-tested-in-Emlen-funnels-can}
  {\bibfield  {journal} {\bibinfo  {journal} {J. Exp. Biol.}\ }\textbf
  {\bibinfo {volume} {218}},\ \bibinfo {pages} {206} (\bibinfo {year}
  {2015})}\BibitemShut {NoStop}%
\bibitem [{\citenamefont {Chen}\ \emph
  {et~al.}(2024{\natexlab{b}})\citenamefont {Chen}, \citenamefont {Liu},\ and\
  \citenamefont {Cai}}]{chen2024identifying}%
  \BibitemOpen
  \bibfield  {author} {\bibinfo {author} {\bibfnamefont {X.}~\bibnamefont
  {Chen}}, \bibinfo {author} {\bibfnamefont {H.}~\bibnamefont {Liu}},\ and\
  \bibinfo {author} {\bibfnamefont {J.}~\bibnamefont {Cai}},\ }\bibfield
  {title} {\bibinfo {title} {{Identifying a possible mechanism for the quantum
  needle in chemical magnetoreception}},\ }\href
  {https://journals.aps.org/pra/abstract/10.1103/PhysRevA.110.042220}
  {\bibfield  {journal} {\bibinfo  {journal} {Phys. Rev. A}\ }\textbf {\bibinfo
  {volume} {110}},\ \bibinfo {pages} {042220} (\bibinfo {year}
  {2024}{\natexlab{b}})}\BibitemShut {NoStop}%
\bibitem [{\citenamefont {{\AA}kesson}\ \emph {et~al.}(2001)\citenamefont
  {{\AA}kesson}, \citenamefont {Morin}, \citenamefont {Muheim},\ and\
  \citenamefont {Ottosson}}]{aakesson2001avian}%
  \BibitemOpen
  \bibfield  {author} {\bibinfo {author} {\bibfnamefont {S.}~\bibnamefont
  {{\AA}kesson}}, \bibinfo {author} {\bibfnamefont {J.}~\bibnamefont {Morin}},
  \bibinfo {author} {\bibfnamefont {R.}~\bibnamefont {Muheim}},\ and\ \bibinfo
  {author} {\bibfnamefont {U.}~\bibnamefont {Ottosson}},\ }\bibfield  {title}
  {\bibinfo {title} {{Avian orientation at steep angles of inclination:
  experiments with migratory white--crowned sparrows at the magnetic North
  Pole}},\ }\href {http://doi.org/10.1098/rspb.2001.1736} {\bibfield  {journal}
  {\bibinfo  {journal} {Proc. R. Soc. Lond. B}\ }\textbf {\bibinfo {volume}
  {268}},\ \bibinfo {pages} {1907} (\bibinfo {year} {2001})}\BibitemShut
  {NoStop}%
\bibitem [{\citenamefont {Hiscock}\ \emph {et~al.}(2016)\citenamefont
  {Hiscock}, \citenamefont {Worster}, \citenamefont {Kattnig}, \citenamefont
  {Steers}, \citenamefont {Jin}, \citenamefont {Manolopoulos}, \citenamefont
  {Mouritsen},\ and\ \citenamefont {Hore}}]{hiscock2016quantum}%
  \BibitemOpen
  \bibfield  {author} {\bibinfo {author} {\bibfnamefont {H.~G.}\ \bibnamefont
  {Hiscock}}, \bibinfo {author} {\bibfnamefont {S.}~\bibnamefont {Worster}},
  \bibinfo {author} {\bibfnamefont {D.~R.}\ \bibnamefont {Kattnig}}, \bibinfo
  {author} {\bibfnamefont {C.}~\bibnamefont {Steers}}, \bibinfo {author}
  {\bibfnamefont {Y.}~\bibnamefont {Jin}}, \bibinfo {author} {\bibfnamefont
  {D.~E.}\ \bibnamefont {Manolopoulos}}, \bibinfo {author} {\bibfnamefont
  {H.}~\bibnamefont {Mouritsen}},\ and\ \bibinfo {author} {\bibfnamefont
  {P.~J.}\ \bibnamefont {Hore}},\ }\bibfield  {title} {\bibinfo {title} {{The
  quantum needle of the avian magnetic compass}},\ }\href
  {https://www.pnas.org/doi/abs/10.1073/pnas.1600341113} {\bibfield  {journal}
  {\bibinfo  {journal} {Proc. Natl. Acad. Sci. USA}\ }\textbf {\bibinfo
  {volume} {113}},\ \bibinfo {pages} {4634} (\bibinfo {year}
  {2016})}\BibitemShut {NoStop}%
\bibitem [{\citenamefont {Cai}\ and\ \citenamefont
  {Plenio}(2013)}]{cai2013chemical}%
  \BibitemOpen
  \bibfield  {author} {\bibinfo {author} {\bibfnamefont {J.}~\bibnamefont
  {Cai}}\ and\ \bibinfo {author} {\bibfnamefont {M.~B.}\ \bibnamefont
  {Plenio}},\ }\bibfield  {title} {\bibinfo {title} {{Chemical compass model
  for avian magnetoreception as a quantum coherent device}},\ }\href
  {https://journals.aps.org/prl/abstract/10.1103/PhysRevLett.111.230503}
  {\bibfield  {journal} {\bibinfo  {journal} {Phys. Rev. Lett.}\ }\textbf
  {\bibinfo {volume} {111}},\ \bibinfo {pages} {230503} (\bibinfo {year}
  {2013})}\BibitemShut {NoStop}%
\bibitem [{\citenamefont {Vitalis}\ and\ \citenamefont
  {Kominis}(2017)}]{PhysRevA.95.032129}%
  \BibitemOpen
  \bibfield  {author} {\bibinfo {author} {\bibfnamefont {K.~M.}\ \bibnamefont
  {Vitalis}}\ and\ \bibinfo {author} {\bibfnamefont {I.~K.}\ \bibnamefont
  {Kominis}},\ }\bibfield  {title} {\bibinfo {title} {{Quantum-limited
  biochemical magnetometers designed using the Fisher information and quantum
  reaction control}},\ }\href {https://doi.org/10.1103/PhysRevA.95.032129}
  {\bibfield  {journal} {\bibinfo  {journal} {Phys. Rev. A}\ }\textbf {\bibinfo
  {volume} {95}},\ \bibinfo {pages} {032129} (\bibinfo {year}
  {2017})}\BibitemShut {NoStop}%
\bibitem [{\citenamefont {Smith}\ \emph {et~al.}(2024)\citenamefont {Smith},
  \citenamefont {Glatthard}, \citenamefont {Chowdhury},\ and\ \citenamefont
  {Kattnig}}]{smith2024optimality}%
  \BibitemOpen
  \bibfield  {author} {\bibinfo {author} {\bibfnamefont {L.~D.}\ \bibnamefont
  {Smith}}, \bibinfo {author} {\bibfnamefont {J.}~\bibnamefont {Glatthard}},
  \bibinfo {author} {\bibfnamefont {F.~T.}\ \bibnamefont {Chowdhury}},\ and\
  \bibinfo {author} {\bibfnamefont {D.~R.}\ \bibnamefont {Kattnig}},\
  }\bibfield  {title} {\bibinfo {title} {{On the optimality of the radical-pair
  quantum compass}},\ }\href
  {https://iopscience.iop.org/article/10.1088/2058-9565/ad48b4/meta} {\bibfield
   {journal} {\bibinfo  {journal} {Quantum Sci. Technol.}\ }\textbf {\bibinfo
  {volume} {9}},\ \bibinfo {pages} {035041} (\bibinfo {year}
  {2024})}\BibitemShut {NoStop}%
\bibitem [{\citenamefont {Degen}\ \emph {et~al.}(2017)\citenamefont {Degen},
  \citenamefont {Reinhard},\ and\ \citenamefont
  {Cappellaro}}]{degen2017quantum}%
  \BibitemOpen
  \bibfield  {author} {\bibinfo {author} {\bibfnamefont {C.~L.}\ \bibnamefont
  {Degen}}, \bibinfo {author} {\bibfnamefont {F.}~\bibnamefont {Reinhard}},\
  and\ \bibinfo {author} {\bibfnamefont {P.}~\bibnamefont {Cappellaro}},\
  }\bibfield  {title} {\bibinfo {title} {{Quantum sensing}},\ }\href
  {https://journals.aps.org/rmp/abstract/10.1103/RevModPhys.89.035002}
  {\bibfield  {journal} {\bibinfo  {journal} {Rev. Mod. Phys.}\ }\textbf
  {\bibinfo {volume} {89}},\ \bibinfo {pages} {035002} (\bibinfo {year}
  {2017})}\BibitemShut {NoStop}%
\bibitem [{\citenamefont {Schirhagl}\ \emph {et~al.}(2014)\citenamefont
  {Schirhagl}, \citenamefont {Chang}, \citenamefont {Loretz},\ and\
  \citenamefont {Degen}}]{schirhagl2014nitrogen}%
  \BibitemOpen
  \bibfield  {author} {\bibinfo {author} {\bibfnamefont {R.}~\bibnamefont
  {Schirhagl}}, \bibinfo {author} {\bibfnamefont {K.}~\bibnamefont {Chang}},
  \bibinfo {author} {\bibfnamefont {M.}~\bibnamefont {Loretz}},\ and\ \bibinfo
  {author} {\bibfnamefont {C.~L.}\ \bibnamefont {Degen}},\ }\bibfield  {title}
  {\bibinfo {title} {{Nitrogen-vacancy centers in diamond: nanoscale sensors
  for physics and biology}},\ }\href
  {https://www.annualreviews.org/content/journals/10.1146/annurev-physchem-040513-103659}
  {\bibfield  {journal} {\bibinfo  {journal} {Annu. Rev. Phys. Chem.}\ }\textbf
  {\bibinfo {volume} {65}},\ \bibinfo {pages} {83} (\bibinfo {year}
  {2014})}\BibitemShut {NoStop}%
\bibitem [{\citenamefont {Biedermann}\ \emph {et~al.}(2015)\citenamefont
  {Biedermann}, \citenamefont {Wu}, \citenamefont {Deslauriers}, \citenamefont
  {Roy}, \citenamefont {Mahadeswaraswamy},\ and\ \citenamefont
  {Kasevich}}]{biedermann2015testing}%
  \BibitemOpen
  \bibfield  {author} {\bibinfo {author} {\bibfnamefont {G.}~\bibnamefont
  {Biedermann}}, \bibinfo {author} {\bibfnamefont {X.}~\bibnamefont {Wu}},
  \bibinfo {author} {\bibfnamefont {L.}~\bibnamefont {Deslauriers}}, \bibinfo
  {author} {\bibfnamefont {S.}~\bibnamefont {Roy}}, \bibinfo {author}
  {\bibfnamefont {C.}~\bibnamefont {Mahadeswaraswamy}},\ and\ \bibinfo {author}
  {\bibfnamefont {M.}~\bibnamefont {Kasevich}},\ }\bibfield  {title} {\bibinfo
  {title} {{Testing gravity with cold-atom interferometers}},\ }\href
  {https://journals.aps.org/pra/abstract/10.1103/PhysRevA.91.033629} {\bibfield
   {journal} {\bibinfo  {journal} {Phys. Rev. A}\ }\textbf {\bibinfo {volume}
  {91}},\ \bibinfo {pages} {033629} (\bibinfo {year} {2015})}\BibitemShut
  {NoStop}%
\bibitem [{\citenamefont {Xiao}\ \emph {et~al.}(2020)\citenamefont {Xiao},
  \citenamefont {Hu}, \citenamefont {Cai},\ and\ \citenamefont
  {Zhao}}]{xiao2020magnetic}%
  \BibitemOpen
  \bibfield  {author} {\bibinfo {author} {\bibfnamefont {D.-W.}\ \bibnamefont
  {Xiao}}, \bibinfo {author} {\bibfnamefont {W.-H.}\ \bibnamefont {Hu}},
  \bibinfo {author} {\bibfnamefont {Y.}~\bibnamefont {Cai}},\ and\ \bibinfo
  {author} {\bibfnamefont {N.}~\bibnamefont {Zhao}},\ }\bibfield  {title}
  {\bibinfo {title} {{Magnetic noise enabled biocompass}},\ }\href
  {https://journals.aps.org/prl/abstract/10.1103/PhysRevLett.124.128101}
  {\bibfield  {journal} {\bibinfo  {journal} {Phys. Rev. Lett.}\ }\textbf
  {\bibinfo {volume} {124}},\ \bibinfo {pages} {128101} (\bibinfo {year}
  {2020})}\BibitemShut {NoStop}%
\bibitem [{\citenamefont {Timmel}\ \emph {et~al.}(1998)\citenamefont {Timmel},
  \citenamefont {Till}, \citenamefont {Brocklehurst}, \citenamefont
  {Mclauchlan},\ and\ \citenamefont {Hore}}]{timmel1998effects}%
  \BibitemOpen
  \bibfield  {author} {\bibinfo {author} {\bibfnamefont {C.~R.}\ \bibnamefont
  {Timmel}}, \bibinfo {author} {\bibfnamefont {U.}~\bibnamefont {Till}},
  \bibinfo {author} {\bibfnamefont {B.}~\bibnamefont {Brocklehurst}}, \bibinfo
  {author} {\bibfnamefont {K.~A.}\ \bibnamefont {Mclauchlan}},\ and\ \bibinfo
  {author} {\bibfnamefont {P.~J.}\ \bibnamefont {Hore}},\ }\bibfield  {title}
  {\bibinfo {title} {{Effects of weak magnetic fields on free radical
  recombination reactions}},\ }\href
  {https://www.tandfonline.com/doi/abs/10.1080/00268979809483134} {\bibfield
  {journal} {\bibinfo  {journal} {Mol. Phys.}\ }\textbf {\bibinfo {volume}
  {95}},\ \bibinfo {pages} {71} (\bibinfo {year} {1998})}\BibitemShut {NoStop}%
\bibitem [{\citenamefont {Ramsey}(1949)}]{ramsey1949new}%
  \BibitemOpen
  \bibfield  {author} {\bibinfo {author} {\bibfnamefont {N.~F.}\ \bibnamefont
  {Ramsey}},\ }\bibfield  {title} {\bibinfo {title} {{A new molecular beam
  resonance method}},\ }\href
  {https://journals.aps.org/pr/abstract/10.1103/PhysRev.76.996} {\bibfield
  {journal} {\bibinfo  {journal} {Phys. Rev.}\ }\textbf {\bibinfo {volume}
  {76}},\ \bibinfo {pages} {996} (\bibinfo {year} {1949})}\BibitemShut
  {NoStop}%
\bibitem [{\citenamefont {Riehle}\ \emph {et~al.}(1991)\citenamefont {Riehle},
  \citenamefont {Kisters}, \citenamefont {Witte}, \citenamefont {Helmcke},\
  and\ \citenamefont {Bord{\'e}}}]{riehle1991optical}%
  \BibitemOpen
  \bibfield  {author} {\bibinfo {author} {\bibfnamefont {F.}~\bibnamefont
  {Riehle}}, \bibinfo {author} {\bibfnamefont {T.}~\bibnamefont {Kisters}},
  \bibinfo {author} {\bibfnamefont {A.}~\bibnamefont {Witte}}, \bibinfo
  {author} {\bibfnamefont {J.}~\bibnamefont {Helmcke}},\ and\ \bibinfo {author}
  {\bibfnamefont {C.~J.}\ \bibnamefont {Bord{\'e}}},\ }\bibfield  {title}
  {\bibinfo {title} {{Optical Ramsey spectroscopy in a rotating frame: Sagnac
  effect in a matter-wave interferometer}},\ }\href
  {https://journals.aps.org/prl/abstract/10.1103/PhysRevLett.67.177} {\bibfield
   {journal} {\bibinfo  {journal} {Phys. Rev. Lett.}\ }\textbf {\bibinfo
  {volume} {67}},\ \bibinfo {pages} {177} (\bibinfo {year} {1991})}\BibitemShut
  {NoStop}%
\bibitem [{\citenamefont {Kaubruegger}\ \emph {et~al.}(2021)\citenamefont
  {Kaubruegger}, \citenamefont {Vasilyev}, \citenamefont {Schulte},
  \citenamefont {Hammerer},\ and\ \citenamefont
  {Zoller}}]{kaubruegger2021quantum}%
  \BibitemOpen
  \bibfield  {author} {\bibinfo {author} {\bibfnamefont {R.}~\bibnamefont
  {Kaubruegger}}, \bibinfo {author} {\bibfnamefont {D.~V.}\ \bibnamefont
  {Vasilyev}}, \bibinfo {author} {\bibfnamefont {M.}~\bibnamefont {Schulte}},
  \bibinfo {author} {\bibfnamefont {K.}~\bibnamefont {Hammerer}},\ and\
  \bibinfo {author} {\bibfnamefont {P.}~\bibnamefont {Zoller}},\ }\bibfield
  {title} {\bibinfo {title} {{Quantum variational optimization of Ramsey
  interferometry and atomic clocks}},\ }\href
  {https://journals.aps.org/prx/abstract/10.1103/PhysRevX.11.041045} {\bibfield
   {journal} {\bibinfo  {journal} {Phys. Rev. X}\ }\textbf {\bibinfo {volume}
  {11}},\ \bibinfo {pages} {041045} (\bibinfo {year} {2021})}\BibitemShut
  {NoStop}%
\bibitem [{\citenamefont {Cai}\ \emph {et~al.}(2012)\citenamefont {Cai},
  \citenamefont {Caruso},\ and\ \citenamefont {Plenio}}]{PhysRevA.85.040304}%
  \BibitemOpen
  \bibfield  {author} {\bibinfo {author} {\bibfnamefont {J.}~\bibnamefont
  {Cai}}, \bibinfo {author} {\bibfnamefont {F.}~\bibnamefont {Caruso}},\ and\
  \bibinfo {author} {\bibfnamefont {M.~B.}\ \bibnamefont {Plenio}},\ }\bibfield
   {title} {\bibinfo {title} {{Quantum limits for the magnetic sensitivity of a
  chemical compass}},\ }\href {https://doi.org/10.1103/PhysRevA.85.040304}
  {\bibfield  {journal} {\bibinfo  {journal} {Phys. Rev. A}\ }\textbf {\bibinfo
  {volume} {85}},\ \bibinfo {pages} {040304} (\bibinfo {year}
  {2012})}\BibitemShut {NoStop}%
\bibitem [{\citenamefont {Hore}\ and\ \citenamefont
  {Mouritsen}(2016)}]{hore2016radical}%
  \BibitemOpen
  \bibfield  {author} {\bibinfo {author} {\bibfnamefont {P.~J.}\ \bibnamefont
  {Hore}}\ and\ \bibinfo {author} {\bibfnamefont {H.}~\bibnamefont
  {Mouritsen}},\ }\bibfield  {title} {\bibinfo {title} {{The radical-pair
  mechanism of magnetoreception}},\ }\href
  {https://www.annualreviews.org/content/journals/10.1146/annurev-biophys-032116-094545}
  {\bibfield  {journal} {\bibinfo  {journal} {Annu. Rev. Biophys.}\ }\textbf
  {\bibinfo {volume} {45}},\ \bibinfo {pages} {299} (\bibinfo {year}
  {2016})}\BibitemShut {NoStop}%
\bibitem [{\citenamefont {Kominis}(2009)}]{kominis2009quantum}%
  \BibitemOpen
  \bibfield  {author} {\bibinfo {author} {\bibfnamefont {I.~K.}\ \bibnamefont
  {Kominis}},\ }\bibfield  {title} {\bibinfo {title} {{Quantum Zeno effect
  explains magnetic-sensitive radical-ion-pair reactions}},\ }\href
  {https://journals.aps.org/pre/abstract/10.1103/PhysRevE.80.056115} {\bibfield
   {journal} {\bibinfo  {journal} {Phys. Rev. E}\ }\textbf {\bibinfo {volume}
  {80}},\ \bibinfo {pages} {056115} (\bibinfo {year} {2009})}\BibitemShut
  {NoStop}%
\bibitem [{\citenamefont {Jones}\ and\ \citenamefont
  {Hore}(2010)}]{jones2010spin}%
  \BibitemOpen
  \bibfield  {author} {\bibinfo {author} {\bibfnamefont {J.~A.}\ \bibnamefont
  {Jones}}\ and\ \bibinfo {author} {\bibfnamefont {P.~J.}\ \bibnamefont
  {Hore}},\ }\bibfield  {title} {\bibinfo {title} {{Spin-selective reactions of
  radical pairs act as quantum measurements}},\ }\href
  {https://www.sciencedirect.com/science/article/pii/S000926141000120X}
  {\bibfield  {journal} {\bibinfo  {journal} {Chem. Phys. Lett.}\ }\textbf
  {\bibinfo {volume} {488}},\ \bibinfo {pages} {90} (\bibinfo {year}
  {2010})}\BibitemShut {NoStop}%
\bibitem [{\citenamefont {{C. R. Timmel and U. Till and B. Brocklehurst and K.
  A. Mclauchlan and P. J. Hore}}(1998)}]{doi:10.1080/00268979809483134}%
  \BibitemOpen
  \bibfield  {author} {\bibinfo {author} {\bibnamefont {{C. R. Timmel and U.
  Till and B. Brocklehurst and K. A. Mclauchlan and P. J. Hore}}},\ }\bibfield
  {title} {\bibinfo {title} {Effects of weak magnetic fields on free radical
  recombination reactions},\ }\href {https://doi.org/10.1080/00268979809483134}
  {\bibfield  {journal} {\bibinfo  {journal} {Mol. Phys.}\ }\textbf {\bibinfo
  {volume} {95}},\ \bibinfo {pages} {71} (\bibinfo {year} {1998})}\BibitemShut
  {NoStop}%
\bibitem [{\citenamefont {Gardiner}\ and\ \citenamefont
  {Zoller}(2004)}]{gardiner2004quantum}%
  \BibitemOpen
  \bibfield  {author} {\bibinfo {author} {\bibfnamefont {C.}~\bibnamefont
  {Gardiner}}\ and\ \bibinfo {author} {\bibfnamefont {P.}~\bibnamefont
  {Zoller}},\ }\href@noop {} {\emph {\bibinfo {title} {Quantum noise: a
  handbook of Markovian and non-Markovian quantum stochastic methods with
  applications to quantum optics}}}\ (\bibinfo  {publisher} {Springer Science
  \& Business Media},\ \bibinfo {year} {2004})\BibitemShut {NoStop}%
\bibitem [{\citenamefont {Huelga}\ \emph {et~al.}(1997)\citenamefont {Huelga},
  \citenamefont {Macchiavello}, \citenamefont {Pellizzari}, \citenamefont
  {Ekert}, \citenamefont {Plenio},\ and\ \citenamefont
  {Cirac}}]{huelga1997improvement}%
  \BibitemOpen
  \bibfield  {author} {\bibinfo {author} {\bibfnamefont {S.~F.}\ \bibnamefont
  {Huelga}}, \bibinfo {author} {\bibfnamefont {C.}~\bibnamefont
  {Macchiavello}}, \bibinfo {author} {\bibfnamefont {T.}~\bibnamefont
  {Pellizzari}}, \bibinfo {author} {\bibfnamefont {A.~K.}\ \bibnamefont
  {Ekert}}, \bibinfo {author} {\bibfnamefont {M.~B.}\ \bibnamefont {Plenio}},\
  and\ \bibinfo {author} {\bibfnamefont {J.~I.}\ \bibnamefont {Cirac}},\
  }\bibfield  {title} {\bibinfo {title} {{Improvement of frequency standards
  with quantum entanglement}},\ }\href
  {https://journals.aps.org/prl/abstract/10.1103/PhysRevLett.79.3865}
  {\bibfield  {journal} {\bibinfo  {journal} {Phys. Rev. Lett.}\ }\textbf
  {\bibinfo {volume} {79}},\ \bibinfo {pages} {3865} (\bibinfo {year}
  {1997})}\BibitemShut {NoStop}%
\bibitem [{\citenamefont {Haberkorn}(1976)}]{haberkorn1976density}%
  \BibitemOpen
  \bibfield  {author} {\bibinfo {author} {\bibfnamefont {R.}~\bibnamefont
  {Haberkorn}},\ }\bibfield  {title} {\bibinfo {title} {{Density matrix
  description of spin-selective radical pair reactions}},\ }\href@noop {}
  {\bibfield  {journal} {\bibinfo  {journal} {Mol. Phys.}\ }\textbf {\bibinfo
  {volume} {32}},\ \bibinfo {pages} {1491} (\bibinfo {year}
  {1976})}\BibitemShut {NoStop}%
\bibitem [{\citenamefont {Steiner}\ and\ \citenamefont
  {Ulrich}(1989)}]{steiner1989magnetic}%
  \BibitemOpen
  \bibfield  {author} {\bibinfo {author} {\bibfnamefont {U.~E.}\ \bibnamefont
  {Steiner}}\ and\ \bibinfo {author} {\bibfnamefont {T.}~\bibnamefont
  {Ulrich}},\ }\bibfield  {title} {\bibinfo {title} {{Magnetic field effects in
  chemical kinetics and related phenomena}},\ }\href
  {https://doi.org/10.1021/cr00091a003} {\bibfield  {journal} {\bibinfo
  {journal} {Chem. Rev.}\ }\textbf {\bibinfo {volume} {89}},\ \bibinfo {pages}
  {51} (\bibinfo {year} {1989})}\BibitemShut {NoStop}%
\bibitem [{\citenamefont {O'Dea}\ \emph {et~al.}(2005)\citenamefont {O'Dea},
  \citenamefont {Curtis}, \citenamefont {Green}, \citenamefont {Timmel},\ and\
  \citenamefont {Hore}}]{o2005influence}%
  \BibitemOpen
  \bibfield  {author} {\bibinfo {author} {\bibfnamefont {A.~R.}\ \bibnamefont
  {O'Dea}}, \bibinfo {author} {\bibfnamefont {A.~F.}\ \bibnamefont {Curtis}},
  \bibinfo {author} {\bibfnamefont {N.~J.}\ \bibnamefont {Green}}, \bibinfo
  {author} {\bibfnamefont {C.~R.}\ \bibnamefont {Timmel}},\ and\ \bibinfo
  {author} {\bibfnamefont {P.}~\bibnamefont {Hore}},\ }\bibfield  {title}
  {\bibinfo {title} {{Influence of dipolar interactions on radical pair
  recombination reactions subject to weak magnetic fields}},\ }\href
  {https://pubs.acs.org/doi/abs/10.1021/jp0456943} {\bibfield  {journal}
  {\bibinfo  {journal} {J. Phys. Chem. A}\ }\textbf {\bibinfo {volume} {109}},\
  \bibinfo {pages} {869} (\bibinfo {year} {2005})}\BibitemShut {NoStop}%
\bibitem [{\citenamefont {Efimova}\ and\ \citenamefont
  {Hore}(2008)}]{efimova2008role}%
  \BibitemOpen
  \bibfield  {author} {\bibinfo {author} {\bibfnamefont {O.}~\bibnamefont
  {Efimova}}\ and\ \bibinfo {author} {\bibfnamefont {P.}~\bibnamefont {Hore}},\
  }\bibfield  {title} {\bibinfo {title} {{Role of exchange and dipolar
  interactions in the radical pair model of the avian magnetic compass}},\
  }\href {https://www.cell.com/fulltext/S0006-3495(08)70595-2} {\bibfield
  {journal} {\bibinfo  {journal} {Biophys. J.}\ }\textbf {\bibinfo {volume}
  {94}},\ \bibinfo {pages} {1565} (\bibinfo {year} {2008})}\BibitemShut
  {NoStop}%
\bibitem [{\citenamefont {Gauger}\ \emph {et~al.}(2011)\citenamefont {Gauger},
  \citenamefont {Rieper}, \citenamefont {Morton}, \citenamefont {Benjamin},\
  and\ \citenamefont {Vedral}}]{PhysRevLett.106.040503}%
  \BibitemOpen
  \bibfield  {author} {\bibinfo {author} {\bibfnamefont {E.~M.}\ \bibnamefont
  {Gauger}}, \bibinfo {author} {\bibfnamefont {E.}~\bibnamefont {Rieper}},
  \bibinfo {author} {\bibfnamefont {J.~J.~L.}\ \bibnamefont {Morton}}, \bibinfo
  {author} {\bibfnamefont {S.~C.}\ \bibnamefont {Benjamin}},\ and\ \bibinfo
  {author} {\bibfnamefont {V.}~\bibnamefont {Vedral}},\ }\bibfield  {title}
  {\bibinfo {title} {{Sustained Quantum Coherence and Entanglement in the Avian
  Compass}},\ }\href {https://doi.org/10.1103/PhysRevLett.106.040503}
  {\bibfield  {journal} {\bibinfo  {journal} {Phys. Rev. Lett.}\ }\textbf
  {\bibinfo {volume} {106}},\ \bibinfo {pages} {040503} (\bibinfo {year}
  {2011})}\BibitemShut {NoStop}%
\bibitem [{\citenamefont {Nielsen}\ and\ \citenamefont
  {Chuang}(2010)}]{nielsen2010quantum}%
  \BibitemOpen
  \bibfield  {author} {\bibinfo {author} {\bibfnamefont {M.~A.}\ \bibnamefont
  {Nielsen}}\ and\ \bibinfo {author} {\bibfnamefont {I.~L.}\ \bibnamefont
  {Chuang}},\ }\href@noop {} {\emph {\bibinfo {title} {{Quantum computation and
  quantum information}}}}\ (\bibinfo  {publisher} {Cambridge university
  press},\ \bibinfo {year} {2010})\BibitemShut {NoStop}%
\end{thebibliography}

\providecommand{\noopsort}[1]{}\providecommand{\singleletter}[1]{#1}%
%


\end{document}